\documentclass[10pt]{wlscirep}
\usepackage[utf8]{inputenc}
\usepackage[T1]{fontenc}  
\usepackage{bm}
\usepackage{siunitx}
\usepackage{xspace}
\usepackage[inkscapelatex=false]{svg}

\usepackage{longtable}
\usepackage{tabularx}
\usepackage{graphicx}
\usepackage{caption}
\usepackage{subcaption}
\usepackage{amsmath}
\usepackage{amssymb}
\usepackage{rotating}
\usepackage{multirow}
\usepackage{afterpage}
\usepackage[labelfont=bf]{caption}
\usepackage{setspace}
\usepackage{enumitem}
\usepackage{upgreek}
\usepackage{float}
\newcommand{\dnn}{DNN\xspace}
\newcommand{\cim}{CIM\xspace}
\newcommand{\cam}{CAM\xspace}
\newcommand{\gap}{GAP\xspace}
\newcommand{\gpu}{GPU\xspace}

\title{In-Memory Implementation of Dynamic Neural Network Using Resistive Memory}
\title{Semantic memory-based dynamic neural network using memristive ternary CIM and CAM for 2D and 3D vision}
\title{Dynamic neural network with memristive CIM and CAM for 2D and 3D vision}

\author[1,3,4]{Yue Zhang}
\author[2,5,7]{Woyu Zhang}
\author[1,3,4]{Shaocong Wang}
\author[1,3,4]{Ning Lin}
\author[1,3,4]{Yifei Yu}
\author[1,3,4]{Yangu He}
\author[1,3,4]{Bo Wang}
\author[6]{Hao Jiang}
\author[8]{Peng Lin}
\author[2,5]{Xiaoxin Xu}
\author[1]{Xiaojuan Qi}
\author[1,3,4,*]{Zhongrui Wang}
\author[6,*]{Xumeng Zhang}
\author[2,5,7,*]{Dashan Shang}
\author[2,6]{Qi Liu}
\author[3,9]{Kwang-Ting Cheng}
\author[2,6]{Ming Liu}
\affil[1]{Department of Electrical and Electronic Engineering, the University of Hong Kong, Hong Kong, China}
\affil[2]{Key Laboratory of Microelectronic Devices \& Integrated Technology, Institute of Microelectronics, Chinese Academy of Sciences, Beijing 100049, China}
\affil[3]{ACCESS – AI Chip Center for Emerging Smart Systems, InnoHK Centers, Hong Kong Science Park, Hong Kong, China}
\affil[4]{Institute of the Mind, the University of Hong Kong, Hong Kong, China}
\affil[5]{Key Lab of Fabrication technologies for Integrated Circuits, Chinese Academy of Sciences, Beijing 100049, China}
\affil[6]{State Key Laboratory of Integrated Chips and Systems, Frontier Institute of Chip and System, Fudan University, Shanghai 200433, China}
\affil[7]{University of Chinese Academy of Sciences, Beijing 100049, China}
\affil[8]{College of Computer Science and Technology, Zhejiang University, Zhejiang, 310027, China}
\affil[9]{Department of Electronic and Computer Engineering, the Hong Kong University of Science and Technology, Hong Kong, China}

\affil[*]{e-mail: zrwang@eee.hku.hk; xumengzhang@fudan.edu.cn; shangdashan@ime.ac.cn}

\begin{abstract}

The brain is dynamic, associative and efficient. It reconfigures by associating the inputs with past experiences, with fused memory and processing. In contrast, AI models are static, unable to associate inputs with past experiences, and run on digital computers with physically separated memory and processing. We propose a hardware-software co-design, a semantic memory-based dynamic neural network (\dnn) using memristor. The network associates incoming data with the past experience stored as semantic vectors. The network and the semantic memory are physically implemented on noise-robust ternary memristor-based Computing-In-Memory (\cim) and Content-Addressable Memory (\cam) circuits, respectively. We validate our co-designs, using a 40nm memristor macro, on ResNet and PointNet++ for classifying images and 3D points from the MNIST and ModelNet datasets, which not only achieves accuracy on par with software but also a 48.1\% and 15.9\% reduction in computational budget. Moreover, it delivers a 77.6\% and 93.3\% reduction in energy consumption.

\end{abstract}

\begin{document}

\flushbottom
\maketitle

\thispagestyle{empty}

\section*{Introduction}


The human brain's efficiency in executing complex tasks on a remarkably low energy budget relies on the synergy of dynamic reconfigurability, associative memory with past experience, and collocation of memory and information processing.

\begin{figure}[!t]
\centering
\includegraphics[width=1.0\linewidth]{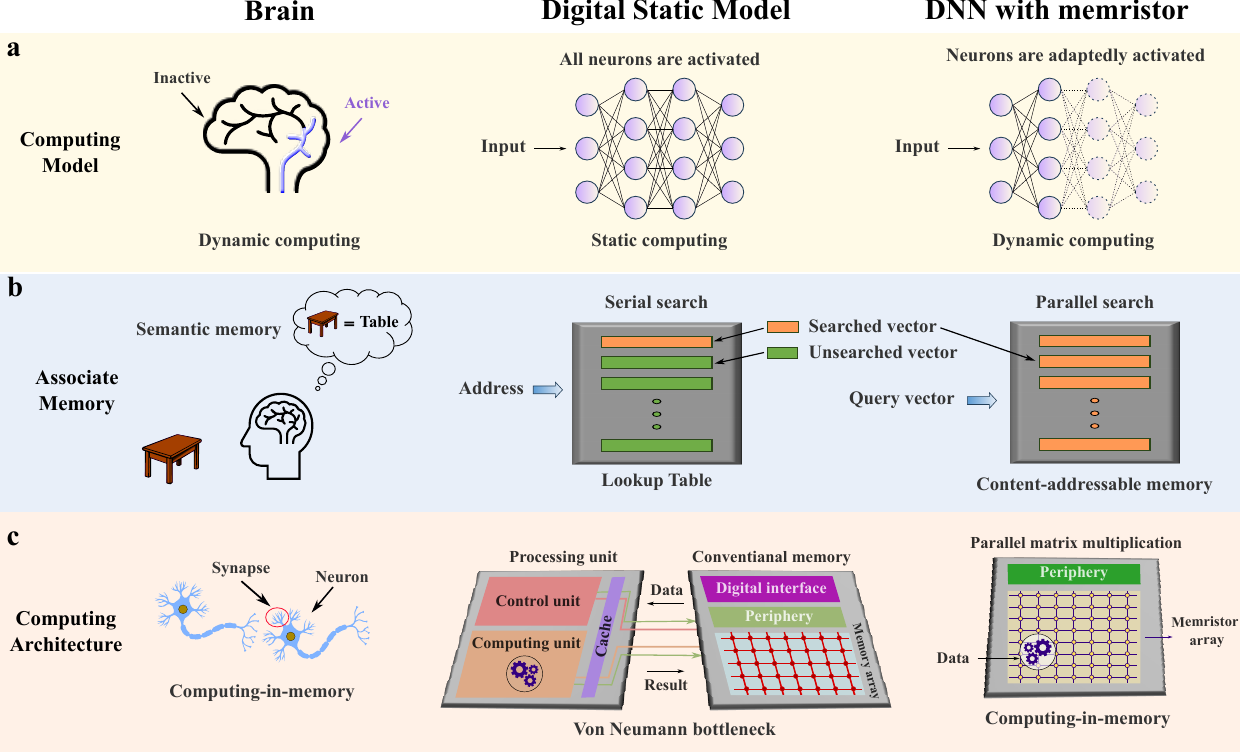}
\caption{\textbf{Brain-inspired dynamic neural network with memristors.} 
\textbf{a,} Comparison of the computing model of the brain, static network and dynamic network. 
\textbf{b,} Comparison of the associative memory mechanism in the brain, digital hardware and memristor-based CAM.
\textbf{c,} Comparison of the computing architecture of the brain, digital hardware, and memristor-based CIM.
}
\label{figintro}
\end{figure}

The brain's processing of information is dynamic and adaptive. Rather than maintaining a static computational model, the brain's neural networks exhibit complex and dynamic activation and connectivity patterns in response to various stimuli and tasks. The dynamic connectivity of the brain primarily hinges on structural plasticity~\cite{hubel1962receptive,murata2000selectivity,diano2017dynamic}. Moreover, the brain displays another layer of dynamic behavior; when individuals encounter objects or tasks of varying complexity, different brain regions are activated~\cite{cichy2016similarity}. Neural activity patterns of the brain can vary across different brain regions and change over time, enabling the brain to integrate sensory information, make decisions, and adapt to new situations~\cite{hermans2014dynamic}. The dynamic nature of the brain allows it to process tasks efficiently while consuming minimal power. In contrast, the majority of neural networks are static with their fixed topology, lacking such adaptability, leading to inefficient resource allocation and limitations in handling evolving information or tasks (Fig.\ref{figintro}(a)).

In addition, the brain is capable of associating new information with its past experience, such as in the realm of semantic and high-level visual processing. Biological experiments~\cite{scarborough1979accessing} have demonstrated the brain's aptitude for accelerating the recognition of familiar objects via memory encoding and retrieval processes within activated neurons, functioning as a semantic cache that allows for rapid retrieval of past experience. Traditional computers, however, rely on address-based information storage and search, rather than associate observations in terms of similarity (Fig.\ref{figintro}(b)).

Moreover, the brain employs chemical synapses not only to store information but also to process information right where they are stored, featuring low power consumption and high parallelism. This is contrasted with the high energy consumption and latency challenges posed by the current von Neumann architecture digital computers, where the separation between computing and storage units necessitates frequent data transfers ~\cite{wong2015memory} (Fig.\ref{figintro}(c)).

Here we propose a hardware-software co-design, semantic memory-based dynamic neural network (\dnn) using memristor, to mirror the 3 characteristics of the brain computing paradigm.

Software-wise, semantic memory-based \dnn aims to equip artificial neural networks with the dynamic reconfigurability of the brain~\cite{han2021dynamic}(Fig.\ref{figintro}(a)). The network associates the new information with its past experience and allocates computations based on demand, more efficient and effective than static networks in many applications~\cite{hu2023dynamic,li2023unleashing,ye2023efficient,yang2022progressive}. Also, the network's adaptability allows it to balance accuracy and efficiency by adjusting to varying computational budgets in real-time~\cite{panda2017energy,teerapittayanon2016branchynet,ju2021learning,park2023dynamic,huang2017multi,shazeer2017outrageously}, for scenarios where computational budgets may change or vary, such as in resource-constrained devices or distributed computing environments ~\cite{tan2019mnasnet}. 


Hardware-wise, we employ memristor, an emerging device that physically resembles the synapses of the brain, to physically implement the ternary weights of noise-robust dynamic neural network and the associative semantic memory, using computing-in-memory (\cim) and content-addressable memory (\cam) circuits, respectively. 

Memristor-based \cim circuits use simple physical laws for matrix-vector multiplications. As matrix weights are physically stored in a memristor array and multiplication is performed right at where the weight is stored, they amalgamate computing and memory elements, which not only overcomes the von Neumann bottleneck but also leads to high parallelism ~\cite{alibart2013pattern,prezioso2015training,yu2016binary,yao2017face,sheridan2017sparse,du2017reservoir,bayat2018implementation,hu2018memristor,moon2019temporal,cai2019fully,duan2020spiking,joshi2020accurate,yao2020fully,xue2021cmos,karunaratne2020memory,liu2020neural,karunaratne2021robust,zhong2021dynamic,milano2022materia,dalgaty2021situ,wang2023echo} (Fig.\ref{figintro}(c)). 

Memristor-based \cam circuit is an associative memory inspired by the brain (Fig.\ref{figintro}(b)), functioning as semantic cache memories in the dynamic neural networks~\cite{yang2019ternary}. Unlike traditional electronic memory that relies on addresses for information retrieval, the brain and \cam feature for parallel content-based searches~\cite{li2020analog,pedretti2022differentiable,laguna2022hardware,mao2022experimentally,pedretti2021tree,ni2019ferroelectric,karunaratne2021robust}. Specifically, \cam measures the distances between input vector and stored vectors within the memory, eliminating the need for data movement like the \cim~\cite{laguna2021memory,9937772,youn2023memristor,ni2019ferroelectric, chang20173t1r,pagiamtzis2006content}. 

In this article, we validate our co-design using a 40nm memristor macro on representative image and 3D points classification tasks. Using ResNet~\cite{he2016deep} and PointNet++~\cite{qi2017pointnet++}, we showcase notable reductions in computational budget, achieving a 48.1\% reduction for classifying the MNIST~\cite{lecun1998gradient} dataset and a 15.9\% reduction for the ModelNet~\cite{wu20153d} dataset by semantic memory-based \dnn. Moreover, we observe substantial improvements in energy efficiency, resulting in a 77.6\% and 93.3\% reduction in energy consumption for the classification of respective datasets when compared to a state-of-the-art graphic processing unit (\gpu). These findings open up avenues for future research on emulating the brain's adaptive allocation of computational resources, associative recall of past experience, and collocation of memory and processing.

\section*{Results}
\subsection*{Hardware–software co-design: Semantic memory-based DNN using memristor-based CIM and CAM} 
\begin{figure}[!t]
\centering
\includegraphics[width=0.9\linewidth]{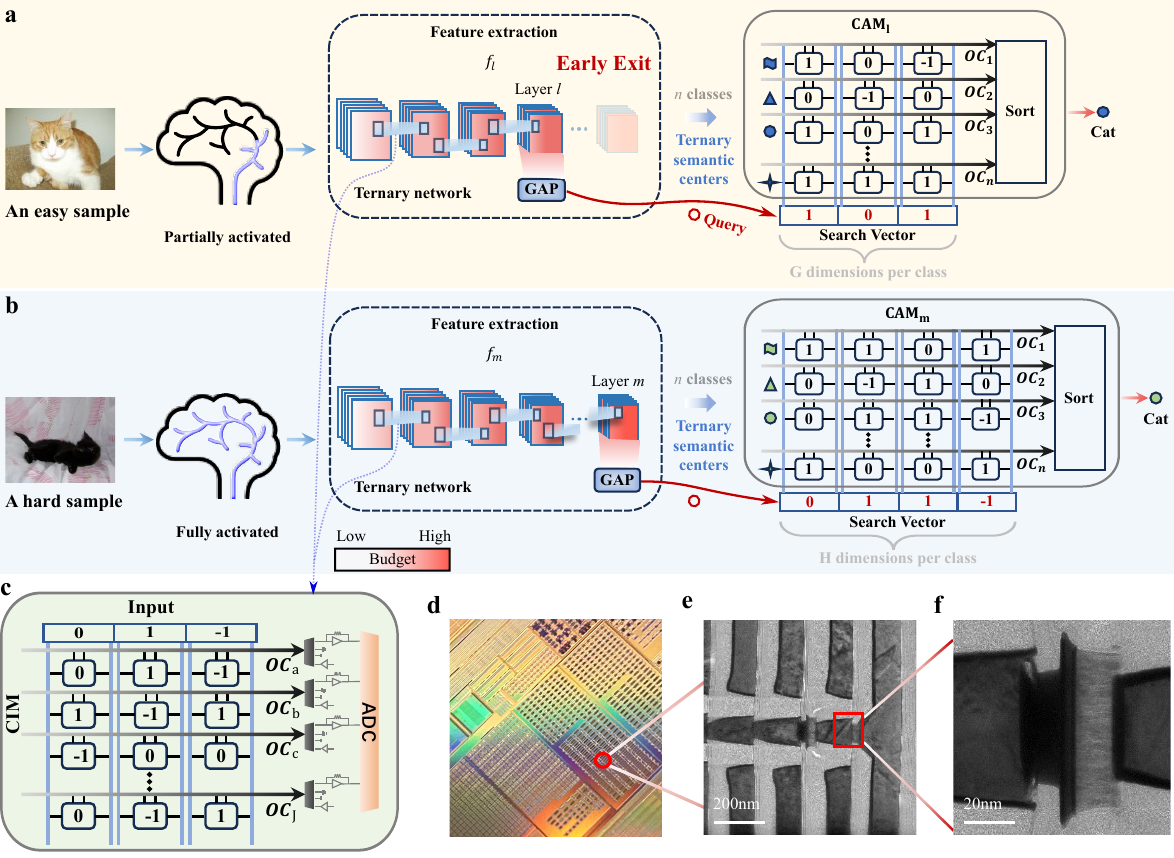}
\caption{\textbf{Hardware-software co-design: Semantic memory-based dynamic neural network using \cim and \cam.} 
\textbf{a,} The proposed architecture consists of a ternary quantized neural network implemented on memristor-based CIM for feature extraction, and an associative memory on memristor-based \cam. Based on the global average pooling (GAP), the network encodes extracted feature maps to low-dimensional ternary semantic center in memory. When a new sample is subsequently queried, the network calculates the semantic vector on each layer’s output feature map and matches them with the cached semantic centers of  \textit{n} classes in memory. The search vector finds the semantic center stored in \cam with maximum cosine similarity, which can be used to predict the class of the query. Once well matched, the network skips the rest of the layers and directly outputs the final results. In this example, an easy sample of a cat can be well classified through early layers with few computing resources and the subsequent layers will not participate in the inference. 
\textbf{b,} If the input is a hard sample, and the early layers fail to provide reliable predictions, classification for this sample requires a deeper network with more computation. 
\textbf{c,} The feature extraction using a memristor-based \cim circuit.
\textbf{d,} Optical photo of the memristor chip.
\textbf{e,} A cross-sectional transmission electron micrograph showcases the memristor crossbar array, fabricated by the backend-of-line process on a 40 nm technology node tape-out.
\textbf{f,} The cross-sectional transmission electron micrograph reveals a solitary nanoscale memristor, operating on the formation and rupture of conducting filaments.
}
\label{fig1}
\end{figure}

Fig.~\ref{fig1} illustrates the semantic memory-based \dnn using memristor-based \cim and \cam. The \dnn optimizes resource allocation and speeds up the forward inference with an early-exiting mechanism according to the semantic memory~\cite{li2021boosting}. The early-exiting dynamically adjusts the number of layers per forward inference, contingent upon the complexity of the input sample, thereby creating a more adaptable network.

As illustrated in Fig.\ref{fig1}(a-b), we use ternary weights (-1, 0, 1) in the network instead of full-precision floating-point numbers (as to be discussed later, ternary quantization outperforms full-precision weights subject to analogue computing noise). Our work uses pre-trained backbone network without training the exits, and can exit from any of the layers before the final layer during inference~\cite{li2021boosting} (see Supplementary Note 1 and Supplementary Table 1 for comparison of early exit work). To compute semantic memory, we infer samples from the training set and apply global average pooling (\gap) to the intermediate layers, converting the feature maps into one-dimensional semantic vectors (see Supplementary Note 2 for global average pooling). These semantic vectors represent the semantic centers for each respective category. Semantic centers undergo ternary quantization and are stored in the \cam. Upon the query of a new test-set sample, the network calculates associated search vectors according to each layer's output feature map and compares them to the cached semantic centers in the \cam (see Supplementary Note 3 for operation of \cam). As such, each semantic memory is able to make a classification based on the cosine similarity between a search vector and the stored semantic centers, at different confidence. 

For relatively simple input samples, the brain can react quickly without the need for deep-level thinking. Similarly, as shown in Fig.\ref{fig1}(a), the forward propagation terminates at a shallow layer (e.g. layer  \textit{l}) with a clearly recognizable cat image. This is because the semantic vector of layer  \textit{l}, when compared to the semantic center vectors in the \cam, yields a sufficiently large confidence in its prediction. The network thus classifies the input accordingly, bypassing the remaining layers and thus enhancing efficiency.
For more complex input samples, as shown in Fig.\ref{fig1}(b), the forward pass goes to deeper layers to capture high-level features for accurate classification. As depicted in the Fig.\ref{fig1}(a-b), the computational budget increases with the number of layers in the forward inference, and contributes to the overall computational cost. Therefore, this early exit approach improves efficiency by avoiding unnecessary computations and streamlining the inference process.

We physically implement the \dnn and semantic memory using a memristor macro. Fig.\ref{fig1}(d) shows the marco integrated onto a printed circuit board along with a Xilinx ZYNQ system-on-chip, forming a hybrid analogue-digital computing platform (see Supplementary Figure 1 for the system design and photo). The analogue \cim and \cam cores consist of CMOS-compatible nanoscale TaN/$\mathrm{TaO}_{\mathrm{x}}$/Ta/TiN memristors (Fig.\ref{fig1}(e-f)) fabricated using the backend-of-line process on a 40 nm technology node tape-out (Fig.\ref{fig1}d).
As shown in Fig.\ref{fig1}(a-c), the ternary weights of the neural network are encoded into the memristive conductances in \cim, and \cam stores the semantic centers.
The input samples are then quantized and mapped to voltages applied to \cim, with the output currents representing output feature maps. The feature maps are then digitized, activated, and pooled in the digital core before being fed to the next layer. The feature maps also undergo \gap, yielding the search vector. The cosine distances between the search vector (as voltage input to the \cam) and the semantic centers stored in the \cam is revealed by the match line signals(e.g. currents here). The currents are then digitized to determine the confidence level associated with each semantic center.

\subsection*{Dynamic ResNet for 2D vision}
\begin{figure}[!t]
\centering
\includegraphics[width=0.8\linewidth]{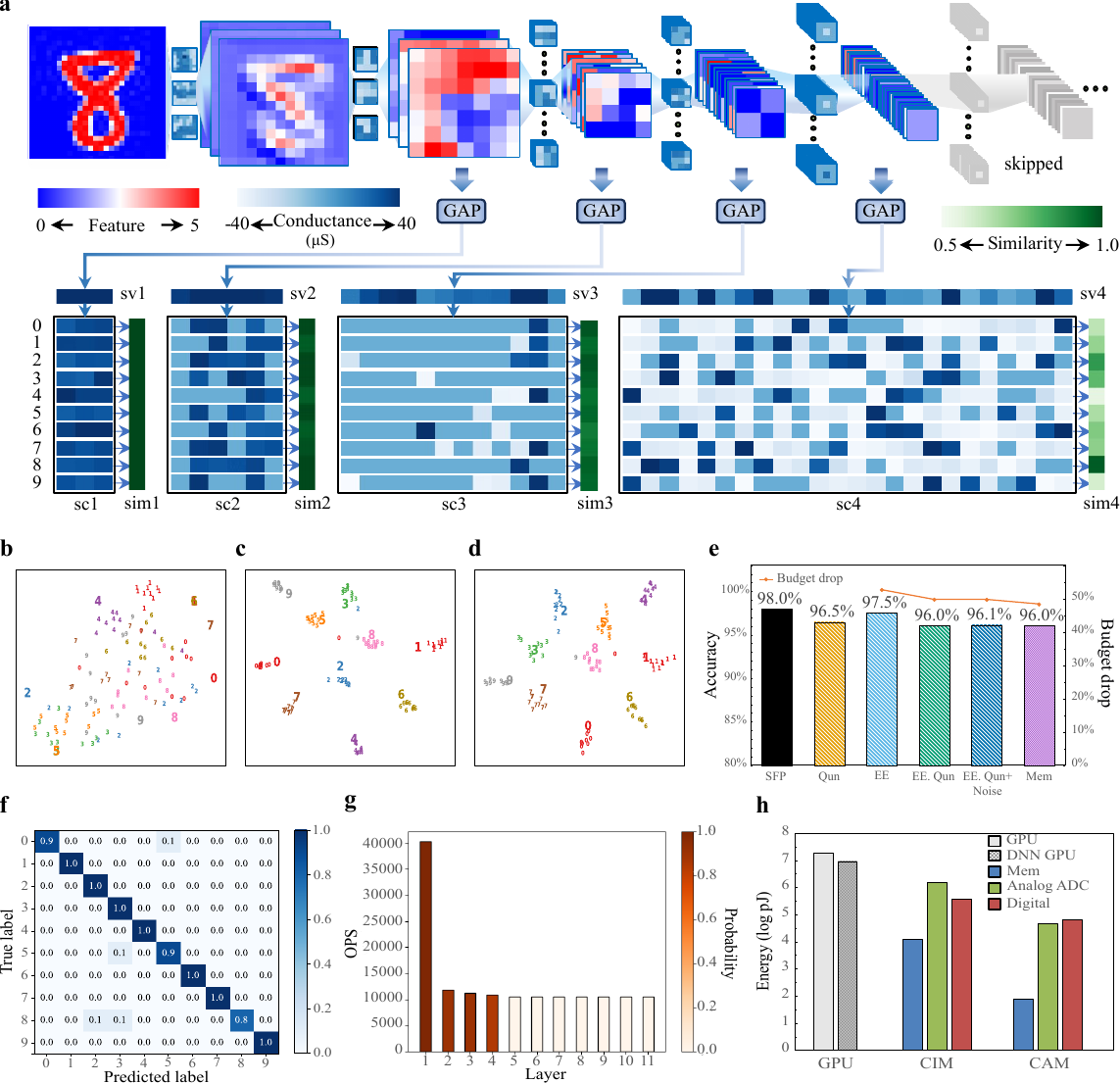}
\caption{\textbf{Hardware-software co-designed dynamic ResNet for MNIST dataset classification.} 
\textbf{a,} Schematic representation of \cim and \cam based dynamic neural network on ResNet. The network extracts feature maps of each residual block, then the feature maps are encoded into semantic vectors (svs) through Global Average Pooling (GAP). Semantic vectors work as keys to lookup corresponding semantic center (sc) by measuring cosine similarities (sim).
\textbf{b-d,} Visualized distribution of semantic vectors (smaller number) and semantic centers (bigger number) for the second, fifth and ninth residual block.
\textbf{e,} The accuracy and budget drop of different model-quantization-noise combinations for ablation study, including static full-precision (SFP), ternary quantization (Qun), early exit (EE), early exit with ternary quantization (EE.Qun), early exit with ternary quantization and noise (EE.Qun+noise), and memristor-based hardware (Mem) experiment.
\textbf{f,} Normalized confusion matrix of memristor-based hardware experiment.
\textbf{g,} Operations (OPS) per layer and the probability of input passing through of each layer.
\textbf{h,} Comparison of the inference energy consumption across \gpu and memristor-based \cim and \cam in a hybrid analogue–digital system.
}
\label{fig2}
\end{figure}

First, we applied our co-design to classify the MNIST dataset~\cite{lecun1998gradient}using the ResNet model~\cite{he2016deep}. Fig.\ref{fig2}(a) shows an experimental example of the forward inference on the memristor-based hybrid analogue-digital system. The model and the semantic memory were first ex situ trained, before being quantized and physically mapped to memristor-based \cim and \cam (see Method). During inference, a handwritten image (e.g. digit eight) is mapped to voltages and the \cim physically computes the feature maps of each residual block. These feature maps produce associated search vector of each residual block after \gap, for being compared with semantic centers of the same block in \cam by measuring cosine similarities. Upon the forward pass reaching the fourth block, the calculated similarity for the eighth class surpasses the threshold, indicating a sufficient confidence level to classify the input handwritten digit as eight. The forward pass thus stops here and all remaining residual blocks are skipped. 

To benchmark the classification performance of semantic centers of different residual blocks, intra-class distance and inter-class distance~\cite{schroff2015facenet}are measured, as shown in Fig.\ref{fig2}(b-d). T-distributed stochastic neighbour embedding (t-SNE) dimension reduction~\cite{van2008visualizing} visualizes the distribution of search vectors (small labels, voltages applied to \cam) of 100 randomly selected test samples and semantic centers (large labels, stored in \cam) from the second, fifth, and nineth residual blocks. It is observed that samples possess significantly different sample-center distances, qualifying the demand of a dynamic and adaptive network. In addition, different residual block develops different classification capability, necessitating unique threshold adjustments for each layer to guarantee optimal performance. Here we optimize the thresholds for semantic memory of each residual block with Tree-structured Parzen Estimator (TPE)~\cite{bergstra2011algorithms,bergstra2013making} as to be discussed later.

Fig.\ref{fig2}(e) shows the ablation and comparison studies. Software static full-precision (SFP) and ternary quantized (Qun) models are of accuracy 98.0\% and 96.5\%, respectively, which slightly reduces to 97.5\% (EE) and 96.0\% (EE. Qun) with the dynamic early-exit. Taking into account analogue memristor noise, the simulated ternary quantized dynamic model (EE. Qun+Noise) shows an accuracy 96.1\%, consistent with our experimental observation (Mem) 96.0\%, accompanied with a 48.1\% reduction of computational budget (see Supplementary Note 4 for performance comparison with LeNet). Fig.\ref{fig2}(f) shows the experimentally acquired confusion matrix, in which the prevalence of diagonal elements corroborates the high classification accuracy. Fig.\ref{fig2}(g) shows the computational budget of inference samples at each residual block and the probabilities of samples passing through each block (see Supplementary Note 5 for computational budget breakdown). It is observed that most samples only need to go through the first four residual blocks while difficult samples propagate to deep residual blocks, which significantly reduces the overall computational budget without compromising the performance. 

Fig.\ref{fig2}(h) shows the energy consumption breakdown of the randomly selected 100 inference samples (see Supplementary Note 6 for operating power and speed estimation). The light (dark) grey bar is the estimated energy consumption of a state-of-the-art \gpu running the static (dynamic) ResNet model, consuming \num{1.83e7} pJ (\num{9.19e6} pJ) (see Supplementary Note 7 for energy consumption modelling). The dynamic model saves about 49.8\% energy due to the decreased network connectivity. For the hybrid analogue-digital system, the blue, green and red bars show energy consumption of memristor (\num{1.21e4} pJ for \cim and \num{77.1} pJ for \cam, see Supplementary Table 2 and Table 3 for energy efficiency of \cam and \cim, respectively), analogue to digital conversion (\num{1.57e6} pJ for \cim and  \num{4.55e4} pJ for \cam), and digital peripherals (\num{3.73e5} pJ for activation and pooling, while \num{6.63e4} pJ for sorting). The overall energy consumption of \dnn on the projected memristor-based hybrid analogue–digital system is approximately \num{2.06e6} pJ, a 77.6\% reduction of the energy consumption compared to software dynamic ResNet due to in-memory computing. Furthermore, compared to static ResNet running on \gpu, our co-design achieves a 8.9-fold enhancement in inference energy efficiency (see Supplementary Note 8 for energy breakdown in ResNet). 


\subsection*{Ternary quantization for analogue noise suppression}
Noise suppression is critical for analogue computing with memristor. There are two major noises in our co-design, the write and read noise. The write noise originates from inevitable programming stochasticity with memristor, while the read noise roots on temporal fluctuations of conductance due to the combined effect of programming instability and other electronic noises ~\cite{yeo2019stuck,niu2016geometric,liu2015vortex}.

\begin{figure}[!t]
\centering
\includegraphics[width=0.8\linewidth]{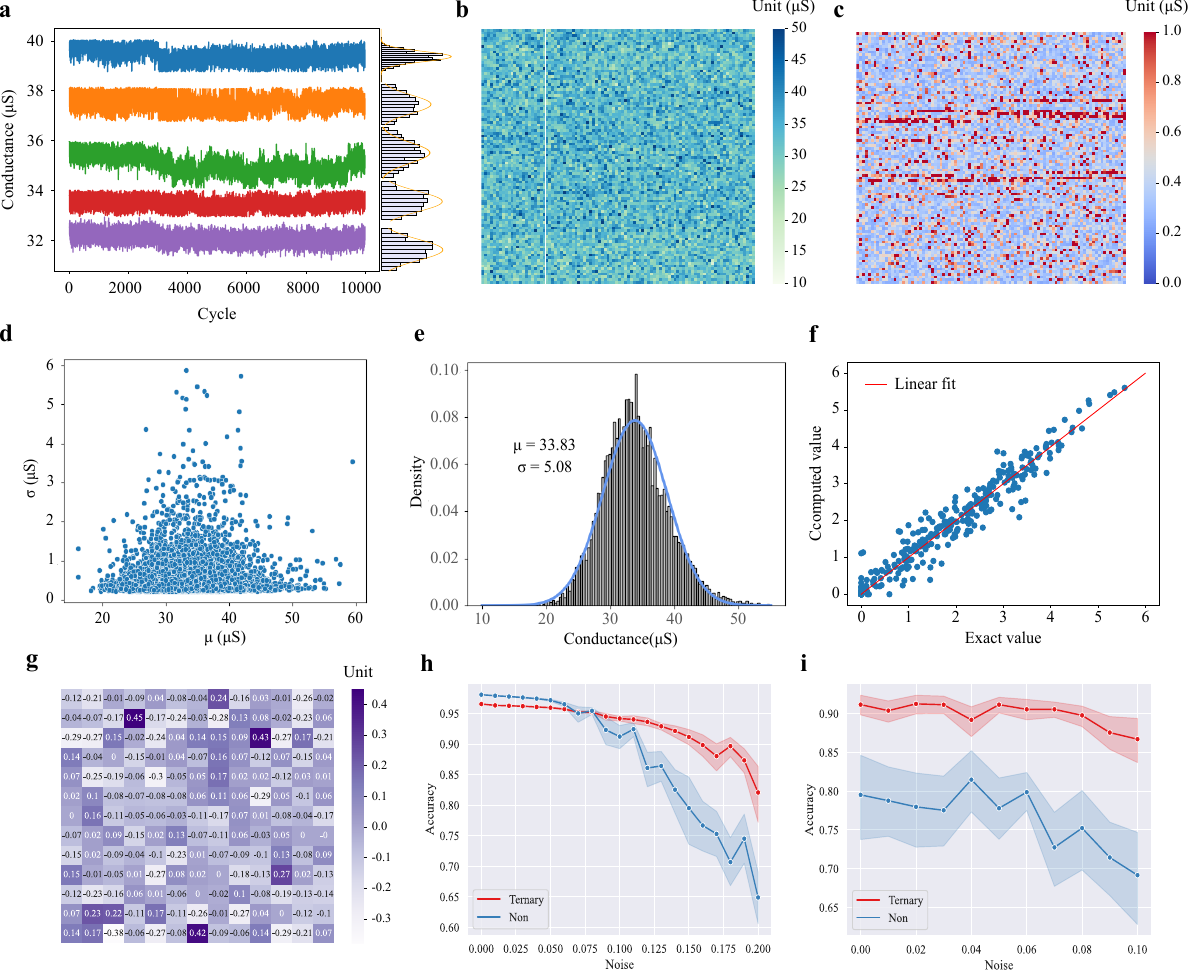}
\caption{\textbf{The intrinsic physical noise of memristor and the mitigation using ternary quantization network} 
\textbf{a,} Conductance variance and statistical histogram of 5 randomly selected memristors with 10,000 read cycles.
\textbf{b-c,} Map of the mean and standard deviation of conductance for 8,930  memristors with 10,000 read cycles. 
\textbf{d,} Scatter plot of the mean and standard deviation of the conductance values of 8,930 memristors. 
\textbf{e,} The histogram of mean conductance in \textbf{b}.
\textbf{f,} Noisy CIM results plotted against noise-free CIM results. The red line represents the ideal case with matched experimental values, while the blue points represent the observed results.
\textbf{g,} Write noise map for the value stored in \cam.
\textbf{h,} Comparison of accuracy with write noise between the non-quantized network and the ternary quantized network.
\textbf{i,} Comparison of accuracy with write and read noise between the non-quantized network and the ternary quantized network.
}
\label{fig3}
\end{figure}

Fig.\ref{fig3}(a) illustrates the two noises. The conductance of five randomly selected memristors programmed under the same condition are continuously sampled 10,000 times. The conductance of each memristor over time follows a quasi-normal distribution, with unique mean and standard deviation. This is also shown in Fig.\ref{fig3}(b), the mean conductance map of a memristor array programmed under the same condition, and Fig.\ref{fig3}(c), the map of conductance standard deviation over 10,000 reading cycles. Fig.\ref{fig3}(d) shows the correlation between mean conductance and the standard deviation. The mean conductance of memristor reflects the write noise, while the standard deviation reflects the read noise. Fig.\ref{fig3}(e) shows the histogram of Fig. \ref{fig3}(b), where the mean conductance of different memristors follows a quasi-normal distribution and ~15\% write noise. The impact of noise on \cim is shown in Fig.\ref{fig3}(f), where final result of the computation is plotted against the exact result.  Such noise also impacts  \cam in a similar manner as examplified by Fig.\ref{fig3}(g).

To mitigate the write and read noise influence over the memristor-based \cim and \cam, we adopt ternary quantization of weights and semantic centers (see Methods), which outperforms mapping full-precision models subject to the write and read noise. We simulate write noise impact on the classification accuracy of dynamic ResNet with and without ternary weight quantization. As shown in Fig.\ref{fig3}(h), the ternary quantized network exhibits considerable resilience against increasing write noise, while the full-precision mapped network performance quickly degrades with increasing programming noise. Furthermore, we simulate read noise impact in conjunction with a fixed 15\% write noise. As shown in Fig.\ref{fig3}(i), the ternary quantized network shows a ~10\% accuracy improvement over direct mapping full-precision weights to memristors.

\subsection*{Dynamic PointNet++ for 3D vision} 
\begin{figure}[!t]
\centering
\includegraphics[width=0.8\linewidth]{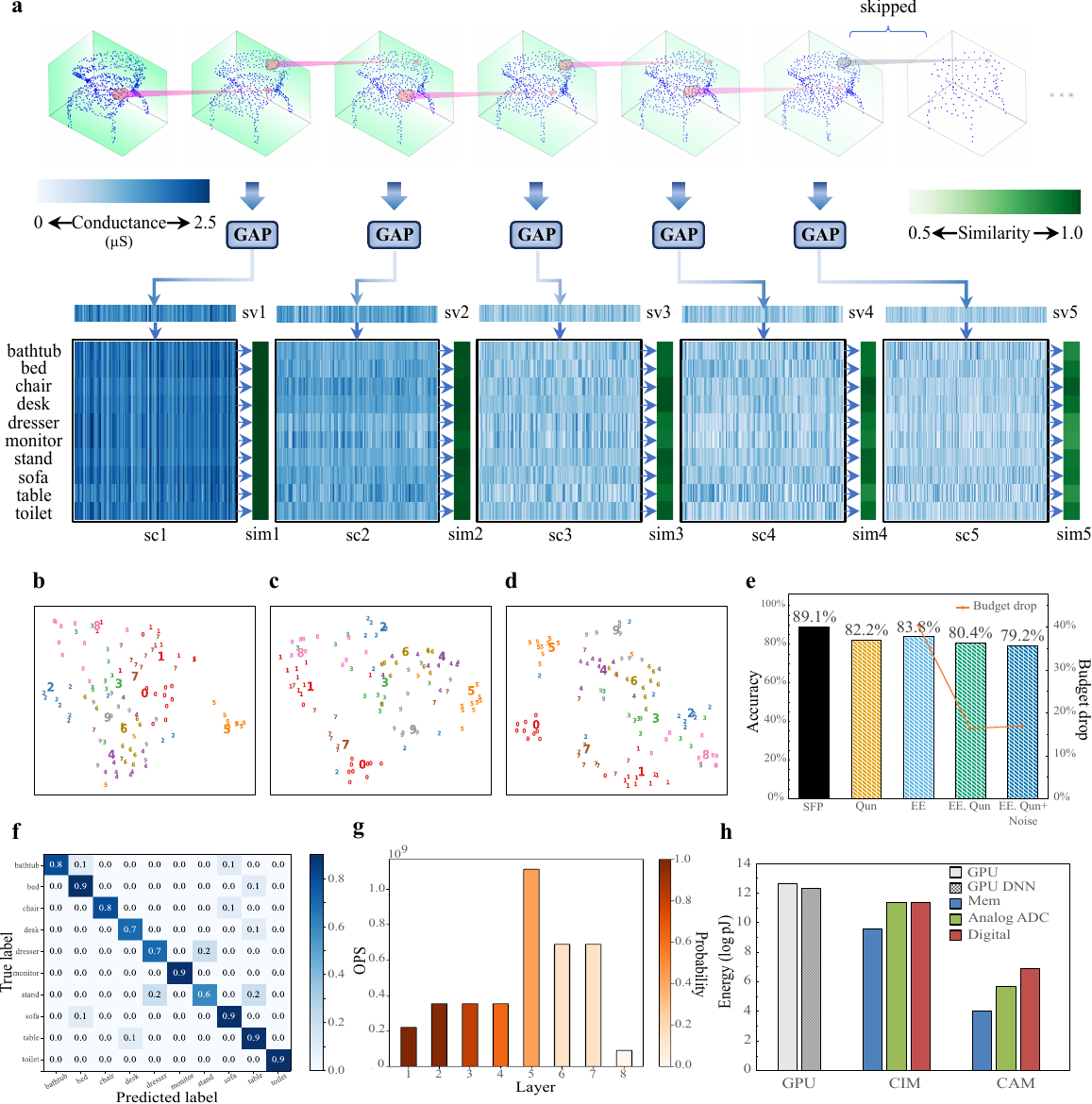}
\caption{\textbf{\cim and \cam based DNN for ModelNet classification using PointNet++.} 
\textbf{a,} Schematic of \cim and \cam based DNN for PointNet++. The input is a 3D point cloud of a chair, the network extracts a 3D feature map of each set abstraction layer, then the feature maps are encoded into semantic vectors (svs) through Global Average Pooling (GAP). Semantic vectors work as keys to lookup corresponding semantic center (sc) by measuring cosine similarities(sim).
\textbf{b-d,} Visualized distribution of semantic vectors (smaller numbers) and semantic centers (bigger numbers) for the second, forth and sixth layer.
\textbf{e,} The accuracy and budget drop of different model-quantization-noise combinations for ablation study, including static full-precision (SFP), ternary quantization (Qun), early exit (EE), early exit with ternary quantization (EE.Qun), early exit with ternary quantization and noise (EE.Qun+noise).
\textbf{f,} Normalized confusion matrix of memristor-based hardware experiment.
\textbf{g,} Operations (OPS) per layer and the probability of input passing through 
 of each layer.
\textbf{h,} Comparison of the inference energy consumption across \gpu and memristor-based \cim and \cam in a hybrid analogue–digital system.
}
\label{fig4}
\end{figure}

Despite 2D images, the advent of 3D sensing technologies has underscored the significance of 3D point clouds in capturing and interpreting the spatial attributes of 3D objects. The classification of 3D point clouds is instrumental in autonomous driving, robotics, and augmented reality. Here, we simulate our co-design using PointNet++~\cite{qi2017pointnet++} model to classify ModelNet~\cite{wu20153d} dataset. The PointNet++ contains multiple set abstraction layers. These layers work in a hierarchical manner, each layer selects representative points, transforms features of all points neighboring to the representative points, and aggregates the neighboring point features to update features of representative points (see Methods for the details of the PointNet++). In this way, the PointNet++ captures both local and global features in a point cloud at different depth of the model, widely used for 3D points classification and segmentation. For better illustration, We randomly select ten categories from the ModelNet dataset. Fig.\ref{fig4}(a) uses a chair as an example, the point features are transformed to voltages and then applied to \cim for feature transformation. The search vectors are calculated for each set abstraction layer and matched with the semantic centers stored in \cam. The system gains enough confidence about the matching results upon forward passing to the fifth set abstraction layer, triggering an early exit and bypassing the remaining layers.

Fig.\ref{fig4}(b-d) uses t-SNE~\cite{van2008visualizing} to visualize the distribution of the search vectors of 100 randomly selected test samples and semantic centers. The second (Fig.\ref{fig4}(b)), the forth (Fig.\ref{fig4}(c)) and the sixth (Fig.\ref{fig4}(d)) set abstraction layer again shows inference samples are of diverse difficulty and the model develops increasing discrimination capability over the depth. 

We also conduct ablation and comparison experiments in Fig.\ref{fig4}(e). The software static model accuracies are 89.1\% and 82.2\% with full-precision and ternary quantized weights, respectively. For dynamic models, the corresponding accuracies are 83.8\% and 80.4\%. The noise slightly reduces the accuracy of dynamic model with ternary weights to 79.2\%, while offering 15.9\% reduction in the computational budget (see Supplementary Note 9 for model size impact on computational saving and accuracy trade-off). Fig.\ref{fig4}(f) is the simulated confusion matrix. The accuracy of categories marked as 3, 4, and 6 is relatively lower, consistent with Fig.\ref{fig4}(b-d) where the search vectors labeled as 3, 4, and 6 cannot be completely segregated. Fig.\ref{fig4}(g) shows the computational budget of input samples at each layer and the probabilities of samples passing through that layer, where the majority of samples exit early by layer 5 (see Supplementary Note 5 for computational budget breakdown).

Fig.\ref{fig4}(h) shows the estimated energy consumption breakdown in inferring samples from 10 randomly selected classes. Software static and dynamic PointNet++ on a state-of-the-art \gpu (light grey bar) consumes \num{4.34e12} pJ and \num{3.65e12} pJ, respectively. The reduction in power consumption can be attributed to decreased network connections and a lower number of operations. For memristor-based hybrid analogue-digital system, the projected energy consumption of memristor, analogue-digital conversion, and digital peripherals are \num{6.35e9} pJ, \num{1.34e11} pJ, and \num{1.53e11} pJ (\num{2.67e4} pJ, \num{7.03e5} pJ, and \num{1.97e7} pJ) for the \cim (\cam). The overall energy consumption of our co-design is approximately \num{2.90e11} pJ, which leads to a 93.3\% reduction when compared to the static PointNet++ in software. The reduction is due to both in-memory computing and DNN (see Supplementary Note 8 for energy breakdown in PointNet++).

\subsection*{Solving optimization problem on trading off between budget and accuracy using TPE}
The threshold of each layer plays a critical role in determining whether an input sample can exit early for computational budget reduction, or keep on forward passing for better accuracy, which balance between accuracy and efficiency. The objective here is to find a Pareto-optimal solution that strikes a trade-off between computational budget and accuracy for the \dnn. We first use grid search to explore the correlation between computational budget and accuracy. Then we design a target function that incorporates both computational budget and accuracy, and use the method of TPE~\cite{bergstra2011algorithms,bergstra2013making} to optimize the threshold of each layer. The optimal solution for the target function yields a \dnn that balances computational budget and accuracy. 

\begin{figure}[!t]
\centering
\includegraphics[width=0.8\linewidth]{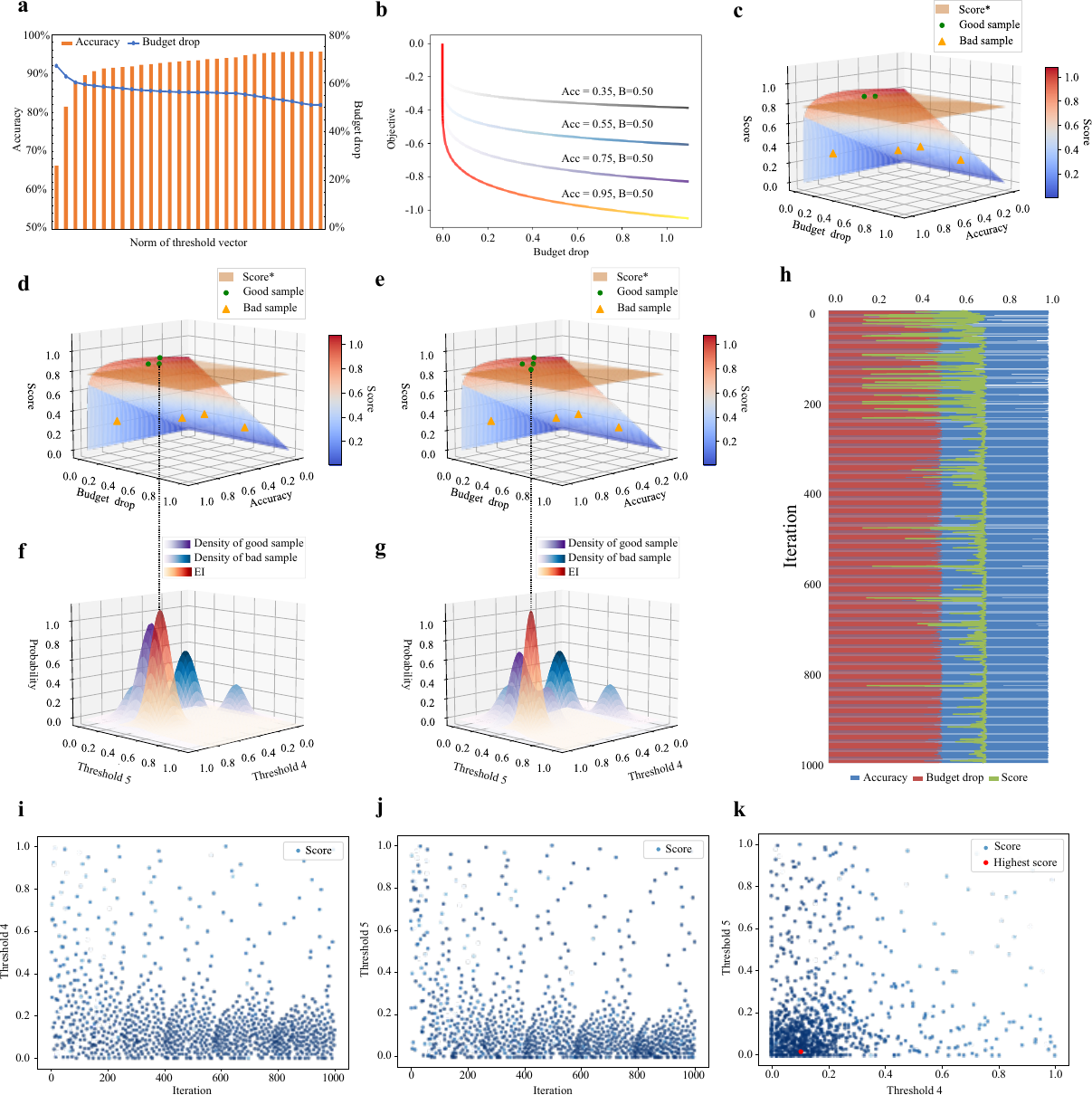}
\caption{\textbf{Solving the optimal problem on trading off between computational budget and accuracy with TPE} 
\textbf{a,} Results of accuracy and budget drop by adjusting the threshold from a lower value to a higher value with the method of grid search.
\textbf{b,} Objective function. Assuming a budget drop of 0.5, and considering accuracy of 0.35, 0.55, 0.75, and 0.95 from top to bottom.
\textbf{c,} Objective score function with accuracy and budget drop. The curved surface represents the score of objective function. The orange surface represents the metric score "Score*", samples with a score higher than Score* are categorized as good samples, samples with a score lower than Score* are categorized as bad samples.
\textbf{d,} The first new sample is obtained by finding thresholds corresponding to the maximum probability calculated by the EI function in figure \textbf{f}.
\textbf{e,} The second new sample is obtained by finding thresholds corresponding to the maximum probability calculated by the EI function in figure \textbf{g}.
\textbf{f,} Method of tree-structured Bayesian optimization. The purple and blue 3D Gaussian surface represent the density of good samples and bad samples in figure \textbf{c}, respectively. The orange surface represents the EI function calculated with the two density functions. 
\textbf{g,} Obtained density of good samples and bad samples in figure \textbf{d} and EI function calculated with the two density functions.
\textbf{h,} Results of accuracy and budget drop with the iteration of TPE.
\textbf{i,} Results of threshold 4 with 1000 iteration times. The color of points represents the value of score, the bluer the color, the higher the score.
\textbf{j,} Results of threshold 5 with 1000 iteration times.
\textbf{k,} Results of obtained score using the searched threshold with 1000 iteration times.
}
\label{fig5}
\end{figure}

Fig.\ref{fig5}(a) shows the grid search results for the thresholds for the co-design using ResNet, which illustrates the trade-off between computational budget and accuracy. Based on that, we devise a target function depends on both computational budget and accuracy.

\begin{equation}
\label{eq1}
\max_{\mathrm{dm} }\quad \mathrm{Acc\left ( dm \right )}  \times \mathrm{\left [\frac{DCB}{B } \right ]} ^{\omega}
\end{equation}
Here, the target budget drop $\mathrm{B=0.50}$, Acc(dm) is the classification accuracy of the \dnn, DCB is the drop of computational budget. The weight factor, $\omega$, is empirically selected to ensure that Pareto-optimal solutions yield similar rewards under different accuracy-budget trade-offs. We observe in grid search that an additional 1\% increase in accuracy corresponds to an approximate 4.35\% increase in computation budget when accuracy is beyond 94\%. Consequently, $\omega$ is set to 0.127. For convenience, we solve the dual problem. Fig.\ref{fig5}(b) shows the negative objective function which decreases with the budget drop or increase of accuracy (see Method)
, which is also visualized in Fig.\ref{fig5}(c) as a curved surface. 

We minimize the negative objective function by identifying the best thresholds of each layer. As TPE method does not model the interaction between thresholds, we use that to solve the Pareto-optimal problem. The first step of TPE is to initialize sampling through random search. We denote the accuracy and budget drop as $\mathbf{x}$ and the calculated score as $y$, the observed results${\{(\mathbf{x}^{(1)},y^{(1)}),..., (\mathbf{x}^{(k)},y^{(k)}) \}}$ are then divided into two groups: the group with good (poor) performance, marked as green dot (yellow triangle)  in Fig.\ref{fig5}(c)), and the splitting value for dividing the two groups is $score^*$. TPE defines the probability distribution $p(\mathbf{x}|y)$ using the following two probability density functions:
\begin{equation}
\label{eq2}
p(\mathbf{x} |y)=\left\{
\begin{aligned} 
l(\mathbf{x} ) \quad \mathrm{if\ y\ge y^* }  \\
g(\mathbf{x} ) \quad \mathrm{if\ y^*> y}  \\
\end{aligned}
\right.
\end{equation}
Fig.\ref{fig5}(f) shows the density function of good samples (purple 3D surface) and the density function of bad samples (blue 3D surface) which are modeled by Parzen estimators (see Methods for a detailed description)~\cite{bergstra2011algorithms,bergstra2013making}. TPE utilizes the following expected improvement (EI) function as an acquisition function:
\begin{equation}
\label{eq3}
\begin{split}
   \mathrm{EI} _{y^*}(\mathbf{x} )  &=  \int_{-\infty }^{y^*}(y^*-y)p(y|\mathbf{x} )dy \\
     &= \int_{-\infty }^{y^*}(y^*-y)\frac{p(\mathbf{x} |y)p(y)}{p(\mathbf{x} )} dy \\
     &= \frac{\gamma y^*l(\mathbf{x} )-l(\mathbf{x} )\int_{-\infty }^{y^*}p(y)dy}{\gamma l(\mathbf{x} )+(1-\gamma )g(\mathbf{x} ) } \propto(\gamma +\frac{g(\mathbf{x} )}{l(\mathbf{x} )}(1-\gamma ) )^{-1}  
\end{split}
\end{equation}
where $\gamma$ is a hyperparameter. The last expression is obtained through Bayes' rule and is proportional to $l(\mathbf{x})/g(\mathbf{x})$ (see Methods for a detailed derivation of the EI equation). Just like Bayesian optimization~\cite{movckus1975bayesian,mockus1998application,jones1998efficient}, the objective is to find the thresholds which increase the probability of $l(\mathbf{x})$ while decreasing the probability of $g(\mathbf{x})$. The new point is searched with thresholds corresponding to the maximum probability of EI function in Fig.\ref{fig5}(f). As shown in Fig.\ref{fig5}(d), the score of the new point is higher than $Score^*$, so it is marked as a good sample. The new sample in Fig.\ref{fig5}(d) is then used for the next iteration to calculate the new density functions and EI function as shown in Fig.\ref{fig5}(g). Similarly, the new sample will be searched with the thresholds corresponding to the maximum probability of EI, and categorized as a good or bad sample for the subsequent iteration. In this way, the TPE adjusts the size of the parameter search space and finds the global optimal solution in as few iterations as possible.

Fig.\ref{fig5}(h) shows iteration results of TPE. In early iterations, there is considerable fluctuation in both the computational budget and accuracy, as well as a large variation in the objective function. However, as the number of iterations approaches 400, the results become stable, indicating a gradual approach towards the optimal solution of the objective function. This is also shown in Fig.\ref{fig5}(i-j) about the evolution of the thresholds of the fourth and fifth layers. The thresholds of both gradually converge to their optimal values, yielding the optimal solution to the Pareto-optimal problem and the best trade-off between the computational budget and accuracy.

\section*{Discussion}
We demonstrate a semantic memory-based \dnn using memristor-based \cim and \cam. \cim processes data directly within memory, resulting in reduced energy consumption and latency compared to conventional digital computers of von Neumann architecture. Additionally, \cam functions as the semantic memory and associates new inputs with past experience in a brain-like manner. Leveraging both \cim and \cam, we physically implement noise-proof ternary quantized \dnn and optimize the thresholds using TPE by solving the Pareto-optimal problem, which results in dynamic connectivities not only reducing computational budget but also retaining network performance.
We validate our co-designs using a 40nm memristor macro on ResNet and PointNet++ for classifying images and 3D points from the MNIST and ModelNet datasets. Our approach achieves accuracy comparable to software while reducing the computation budget by 48.1\% and 15.9\% compared to static neural networks. Furthermore, it offers a significant reduction in energy consumption, with a 77.6\% and 93.3\% decrease compared to a state-of-the-art \gpu.
Our results lay the foundation for future machine intelligence that can potentially parallel the adaptability and efficiency of the human brain.

\section*{Materials and Methods}
\subsection*{Fabrication of memristor array}
Memristors are integrated on the 40nm standard logic platform, forming a 512$\times$512 crossbar array. The memristors are constructed between the metal 4 and metal 5 layers of the backend-of-line process, including bottom electrodes, top electrodes, and a transition metal oxide dielectric layer. The bottom electrodes have a patterned via with a diameter of 60nm, formed with the method of photolithography and etching. Physical vapor deposition and chemical mechanical polishing are employed to deposit material within the via. A 10 nm TaN buffer layer is deposited on the bottom electrode via using physical vapor deposition. Subsequently, 5 nm Ta is deposited and then oxidized in an oxygen environment to form an 8 nm $\mathrm{Tao_{x } }$  dielectric layer. On the top electrode, we sequentially deposited 3 nm Ta and 40 nm TiN using physical vapor deposition. Next, the logic backend-of-line metal is deposited using standard logic process. Memristors in the same column share top electrode connections, and memristors in the same row share bottom electrode connections. After a post-annealing process in vacuum at 400°C for 30 minutes,the fabrication of the memristor chip is completed.

\subsection*{Hybrid analogue–digital computing platform}
The hybrid analogue–digital computing platform consists of a 40 nm memristor chip and the a Xilinx ZYNQ system-on-chip. There is an 8-channel digital-to-analogue converter (DAC80508, TEXAS INSTRUMENTS) in the system to generate parallel 64-channel analogue voltages which ranges from 0V to 5V, it is used to transform input digital signals into corresponding analogue signals. The convergence currents are converted to voltages using a transimpedance amplifier (OPA4322-Q1, TEXAS INSTRUMENTS) and read out as digital signals by an analogue-to-digital converter (ADS8324, TEXAS INSTRUMENTS) with a 14-bit resolution for signal collections. To perform vector-matrix multiplication, a 4-channel analogue multiplexer (CD4051B, TEXAS INSTRUMENTS) with an 8-bit shift register (SN74HC595, TEXAS INSTRUMENTS) applies DC voltages to the word lines of the memristor chip. The multiplication results carried by the current from the source line are converted into voltages and passed to the Xilinx ZYNQ system-on-chip for further processing.

\subsection*{DNN-based ResNet}
Taking advantage of the high and low resistance states exhibited by memristors, we leverage the properties of two such memristors to effectively represent a ternary value, employing the principles of Ohm's law and Kirchhoff's laws. The memristors in crossbar array are partitioned into two matrices for programming the weights of ResNet and values stored in \cam. If both corresponding memristors are adjusted to the high-resistance state, they represent a weight of 0. If the memristor in the first part is in the low resistance state while the second memristor is in the high resistance state, they represent a weight of 1. Conversely, if the memristor in the first part is in the high resistance state while the second memristor is in the low resistance state, they represent a weight of -1. In the experiment, the ResNet model consists of 11 residual blocks, with a total of approximately 88k weight parameters, and there are approximately 2k values stored in \cam.

\subsection*{DNN-based PointNet++}
PointNet++ is built upon the foundation of PointNet, which consists of a series of Multi-Layer Perceptrons (MLPs) and a global feature aggregation step using a symmetric function. PointNet++ uses Set Abstraction (SA) layers to capture local features in the point cloud using Farthest Point Sampling (FPS) strategy to select representative points and apply local PointNet operations on their neighboring points within a specific radius. In the experiment, the DNN-based PointNet++ consists of eight SA layers with varying radius and numbers of representative points. Additionally, PointNet++ consists of Feature Propagation (FP) Layers, which use nearest neighbor interpolation and feature concatenation from different layers to help the network reconstruct global point cloud features from the hierarchical local features. 

\subsection*{Ternary Quantization}
When training a ternary network, ternary quantization is performed during the forward propagation, while weight adjustment using full-precision values is carried out during the backward propagation. The method of ternary quantization is as follows:
\begin{equation} 
\label{ternary}
\begin{split}
w_{min}&=argmin\ W^{l},    \\
w_{max}&=argmax\ W^{l},      \\
l_{in}&=w_{min}+(w_{max}-w_{min})/3,\\
h_{in}&=w_{max}-(w_{max}-w_{min})/3,\\
  \end{split}
\end{equation}
where $W^{l}$ represent the whole weight of block l, $l_{in}$ and $h_{in}$ are two intervals. The output ternary quantized weight is as follows:
\begin{equation} 
\label{ternary}
\begin{split}
w_{q}=\left\{
\begin{aligned}
-1&, \ if\ w_{i}<l_{in},\\
0&, \ if\ h_{in}\ge w_{i}\ge l_{in},\\
1&, \ if\ w_{i}>h_{in},\\
\end{aligned}
\right.
  \end{split}
\end{equation}
\subsection*{Tree Parzen Estimators}

Bayes' rule is a fundamental concept in probability theory and statistics. It describes the probability of an event based on prior knowledge or information. The formula for Bayes' theorem is expressed as:

\begin{equation}
\label{eq4}
\mathrm{P(A|B)=\frac{P(B|A)P(A)}{P(B)}} 
\end{equation}

Where P(A|B) represents the conditional probability of event A given event B. P(B|A) represents the conditional probability of event B given event A. P(A) is the probability of event A occurring. P(B) is the probability of event B occurring. The advantage of using Bayesian inference is that it leverages prior experience to make inferences about the next sample. By incorporating prior information into the analysis, Bayesian methods can accelerate the process of finding the optimal solution. TPE uses EI as the acquisition function. However, since p(y|x) cannot be obtained directly, we employ Bayes' rule of ~\ref{eq4} to perform the transformation in ~\ref{eq3}.
The details for deriving the final expression in ~\ref{eq3} are as the following:
\begin{equation}
\label{eq5}
p(\mathbf{x} )=\int p(\mathbf{x} |y)p(y)dy=\gamma l(\mathbf{x} )+(1-\gamma )g(\mathbf{x} )
\end{equation}

\begin{equation}
\label{eq6}
\int_{-\infty }^{y^*} (y^*-y)p(\mathbf{x} |y)p(y)dy=l(\mathbf{x} )\int_{-\infty }^{y^*}(y^*-y)p(y)dy-l(\mathbf{x} )\int_{-\infty }^{y^*}p(y)dy 
\end{equation}
Where $\gamma$ represents the quantile of TPE, used to divide $l(\mathbf{x})$ and $g(\mathbf{x})$, ranging between 0 and 1. For example, if $\gamma$ is set to 0.2 and there are 10 observed samples, the top 2 best performed samples will be used to construct the distribution of good samples, while the remaining 8 samples will be used to construct the distribution of bad samples.

The the Parzen window~\cite{parzen1962estimation} is a technique for kernel density estimation. It estimates the probability density of an estimated value based on the current observed data and the type of prior distribution. The probability density estimation function for Parzen window is as follows:
\begin{equation}
\label{eq7}
p(\mathbf{x} )=\frac{k}{n\cdot V} =\frac{1}{nh^2}\sum_{i=1}^{n}\phi (\frac{\mathbf{x_i} -\mathbf{x} }{h} )  
\end{equation}
The window function determines weight assigned to each observed data point based on its distance from the point $\mathbf{x_i}$, which means that the weights are higher for data points closer to $\mathbf{x_i}$ and decrease as the distance increases. When using a Gaussian function as the window function, the density function is as:
\begin{flalign}
\label{eq8}
p(\mathbf{x} )=\frac{1}{n}\sum_{i=1}^{n}\frac{1}{\sqrt{2\pi }\sigma  }  exp(-\frac{(\mathbf{x_i} -\mathbf{x} )}{2\sigma ^2} )
\end{flalign}
where $\int p(\mathbf{x} )d\mathbf{x} =1$.

\section*{Acknowledgements}
\subsection*{Funding}
This research is supported by the National Key R\&D Program of China (Grant No. 2020AAA0109005), the National Natural Science Foundation of China (Grant Nos. 62122004, 62374181, 62488201), the Strategic Priority Research Program of the Chinese Academy of Sciences (Grant No. XDB44000000), Beijing Natural Science Foundation (Grant No. Z210006), Hong Kong Research Grant Council (Grant Nos. 27206321, 17205922, 17212923). This research is also partially supported by ACCESS – AI Chip Center for Emerging Smart Systems, sponsored by Innovation and Technology Fund (ITF), Hong Kong SAR.
\subsection*{Author contributions}
Conceptualization: YZ, ZW, DS. 
Methodology: YZ, ZW, DS, WZ.
Investigation: YZ, DS, ZW.
Visualization: YZ, DS, ZW.
Funding acquisition: HJ, PL, XX, DS, ZW, XZ, QL, KTC, ML.
Project administration: DS, ZW.
Supervision: ZW, DS, XQ.
Writing – original draft: YZ.
Writing – review and editing: ZW, DS, YZ, SW, NL, YY, YH, BW, HJ, PL, XX, XQ, WZ, XZ, QL, KTC, ML.
\subsection*{Competing Interests}
The authors declare no competing interests.
\subsection*{Data and materials availability}
All data needed to evaluate the conclusions in the paper are present in the paper and/or the Supplementary Materials.


\newpage 

\end{document}